\documentclass[11pt,twoside]{article}
\usepackage{asp2010}

\resetcounters

\aspvoltitle{Galaxy Evolution:\ Infrared to Millimeter Wavelength Perspective}
\aspcpryear{2011}

\begin{document}

\title{Constraints on the Star-Forming Interstellar Medium in Galaxies Back to the First Billion Years of Cosmic Time}
\author{Dominik A.~Riechers$^1$
\affil{$^1$California Institute of Technology, 1200 East California Blvd, MC\,249-17, Pasadena, CA 91125, USA}}

\begin{abstract}

Constraints on the molecular gas content of galaxies at high redshift
are crucial to further our understanding of star formation and galaxy
evolution through cosmic times, as molecular gas is the fuel for star
formation. Since its initial detection at large cosmic distances
almost two decades ago, studies of molecular gas in the early universe
have come a long way. We have detected CO emission from $>$100
galaxies, covering a range of galaxy populations at $z$$>$1, reaching
out to $z$$>$6, down to sub-kpc scale resolution, and spanning $\sim$2
orders of magnitude in gas mass (aided by gravitational
lensing). Recently, it has even become possible to directly identify
distant galaxies through their molecular emission lines without prior
knowledge of their redshifts.  The new generation of powerful long
wavelength interferometers such as the Expanded Very Large Array
(EVLA) and Atacama Large (sub)Millimeter Array (ALMA) thus hold the
promise to liberate studies of molecular gas in high redshift galaxies
from their heavy pre-selection. This will enable more systematic
studies of the molecular gas content in star-forming galaxies back to
the earliest cosmic times.

\end{abstract}

\section{Introduction}

Recent years have shown a tremendous progress in our understanding of
star formation and stellar mass assembly in galaxies through cosmic
 times. A key diagnostic in such studies is the so-called star
formation history of the universe (SFHU), which describes the volume
density of star formation in the universe as a function of cosmic time
(Madau et al.\ 1996; Lilly et al.\ 1996). Since the initial studies 15
years ago, it has become possible to study the buildup of stellar mass
through the SFHU as a function of galaxy type and mass, star formation
rate (SFR), and cosmic environment, reaching back to the epoch of
cosmic reionization, within 1\,Gyr of the Big Bang. These studies have
shown that the comoving SFR density increases by more than an order of
magnitude between $z$$\sim$0 and 1, peaks between $z$=1 and 3, and
then likely drops gradually out to $z$$>$6--10 (Hopkins \& Beacom
2006; Bouwens et al.\ 2011). According to this relation, about half of
the stellar mass in spheroidal galaxies in the present day universe is
built up at 1$<$$z$$<$3 (leading to the emergence of the Hubble
sequence), which thus is dubbed the `epoch of galaxy assembly'
(Marchesini et al.\ 2009).  Also, the steep incline toward $z$$\sim$1
further steepens for galaxies with higher SFRs, suggesting a
substantially higher abundance of intensely star-forming galaxies at
high $z$ (Le Floc'h et al.\ 2005; Magnelli et al.\ 2009). On the other
hand, the specific SFR (i.e., SFR per total stellar mass of a galaxy)
suggests that the more massive a galaxy, the more of its stellar mass
it forms at earlier cosmic times, as the most massive galaxies only
show quiescent star formation since $z$$\sim$1 (`downsizing'; e.g.,
Zheng et al.\ 2007). Overall, the build-up of stellar mass follows the
SFR density fairly well (Borch et al.\ 2006; Bell et al.\ 2007).

An important, complementary, yet less systematically used diagnostic
in studies of galaxy evolution is the molecular gas content of
galaxies. Molecular gas is a precursor for star formation, as it is
the material out of which stars form. Thus, the relative amount of
molecular gas in a galaxy indicates its future potential for star
formation, and how much stellar mass ($M_\star$) a galaxy can assemble
by $z$=0 without external gas supply.

This work summarizes the progress in studies of molecular gas at high
redshift since the initial investigations almost 2 decades ago. Based
on recent breakthroughs in these investigations, it is outlined how it
may become possible to study the molecular gas content in distant
galaxies in a more systematic fashion in the future.

\section{Molecular Gas in High Redshift Galaxies:\ Fundamentals}

Molecular gas (CO) has been detected in $>$100 high redshift ($z$$>$1)
galaxies to date (see reviews by Solomon \& Vanden Bout 2005; Omont
2007), out to $z$=6.42 (e.g., Walter et al.\ 2003, 2004; Bertoldi et
al.\ 2003; Riechers et al.\ 2009a). Except for a few strongly lensed
systems (e.g., Baker et al.\ 2004; Coppin et al.\ 2007; Riechers et
al.\ 2010a; Riechers 2011), the bulk of these galaxies have massive
gas reservoirs of $M_{\rm gas}$$>$10$^{10}$\,$M_\odot$, as determined
from their CO line intensities.

A small fraction of these gas-rich galaxies could be spatially
resolved in high-resolution CO observations down to 0.15$''$--0.3$''$
resolution (corresponding to 1--2\,kpc at $z$$>$4; e.g., Carilli et
al.\ 2002; Walter et al.\ 2004; Riechers et al.\ 2008a, 2008b, 2009b;
Tacconi et al.\ 2008; Bothwell et al.\ 2010; Engel et al.\ 2010). Such
studies allow to measure the surface density, dynamics, and overall
morphology of the gas, which provides insight on gas supply mechanisms
and on the processes driving star formation (e.g., major mergers
vs.\ gas dynamics driven by ordered rotation). Moreover, the
constraints on the dynamical structure allow to determine the
dynamical mass $M_{\rm dyn}$, and thus, the shape and depth of the
gravitational potential of a galaxy. As $M_{\rm dyn}$ encompasses all
components in the gas-rich region of a galaxy (i.e., $M_{\rm gas}$,
dust mass, $M_\star$, supermassive black hole mass, and dark matter),
it enables an independent mass calibration of the most massive
constituents, and allows to set upper limits on contributors that are
not directly detectable. This is of particular importance for high-$z$
quasars, where the stellar hosts are not detectable at
optical/infrared wavelengths due to the brilliance of the luminous
active galactic nucleus (e.g., Walter et al.\ 2004; Riechers et
al.\ 2008a, 2008b). It also provides independent limits on dark matter
fractions in galaxies where all baryonic components are individually
detectable, and an independent calibration for the conversion factor
from CO line luminosity to $M_{\rm gas}$ (e.g., Daddi et al.\ 2010a).

A main diagnostic for the physical properties of the gas only becomes
accessible by studying multiple CO lines in a high redshift galaxy
(e.g., Wei\ss\ et al.\ 2005, 2007; Riechers et al.\ 2006a). Due to the
increasing rotational energy required to excite higher-level
rotational transitions, it becomes increasingly difficult to
collisionally excite high-$J$ lines\footnote{$J$ stands for the upper
  level rotational quantum number of a rotational CO $J$=n$\to$n--1
  transition.} high enough to keep them in thermal equilibrium with
lower-$J$ lines. Thus, the relative strength of the emission from
different rotational lines allows to constrain the physical properties
of the gas (for collisional excitation, in particular the gas density
and temperature). This implies that the fundamental CO($J$=1$\to$0)
transition is, in general, a better tracer of the total amount of
molecular gas in a galaxy (and thus, $M_{\rm gas}$), as it is less
sensitive to gas excitation conditions than higher-$J$ lines.

One important aspect to keep in mind is that, due to the low critical
density of the CO($J$=1$\to$0) line ($\sim$300\,cm$^{-3}$), it is a
good tracer of the total amount of molecular gas, but not a particular
tracer of the dense regions where star formation actively takes
place. Such star-forming cores are traced more reliably by molecules
with a higher dipole moment than CO, such as HCN, HCO$^+$, HNC, or
CN. The $J$=1$\to$0 transitions of HCN, HCO$^+$, and HNC have critical
densities in excess of 10$^4$\,cm$^{-3}$, and thus, are good tracers
of dense cores. However, due to substantially lower abundances and the
more stringent excitation conditions, the ground-level transitions of
these molecules typically exhibit 10--30$\times$ lower line fluxes
than CO($J$=1$\to$0) (e.g., Gao et al.\ 2007; Riechers et
al.\ 2007a). Thus, molecules different than CO have only been detected
in a handful of high-$z$ galaxies, and only two systems (the
Cloverleaf at $z$=2.56 and APM\,08279+5255 at $z$=3.91) currently have
measurements in more than a single line of a dense gas tracer (e.g.,
Solomon et al.\ 2003; Vanden Bout et al.\ 2004; Carilli et al.\ 2005;
Wagg et al.\ 2005; Riechers et al.\ 2006b, 2007b, 2010b, 2011;
Garcia-Burillo et al.\ 2006; Wei\ss\ et al.\ 2007; Gao et al.\ 2007;
Guelin et al.\ 2007). Thus, studies of the chemical composition and
excitation of the dense, actively star-forming component in the
molecular interstellar medium of high-$z$ galaxies are still rather
limited.

\section{CO Detections at High Redshift:\ Galaxy Populations}

%%%%%%%%%%%%%%%%%%%%%%%%%%%%%%%%%%%%%%%%%%%%%%%%%%%%%%%%%
%%%% Fig.1: Counts
%%%%%%%%%%%%%%%%%%%%%%%%%%%%%%%%%%%%%%%%%%%%%%%%%%%%%%%%%

\begin{figure}[h!]

\vspace*{-10mm}
      \begin{center}
%\hspace*{-60mm}
\includegraphics[width=12.5cm,angle=-0]{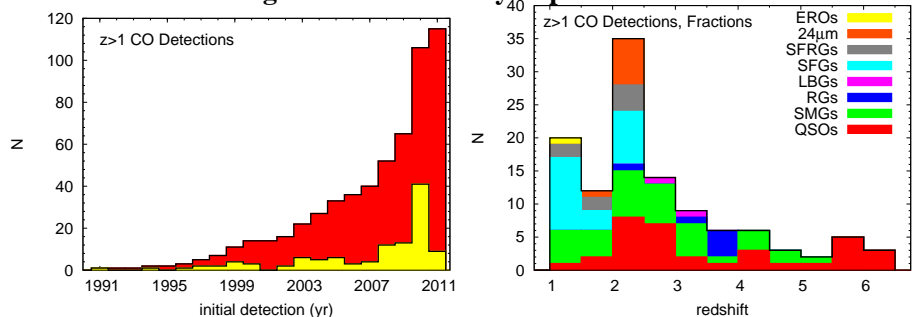}
      \end{center}
\vspace*{-8.5mm}

\caption{Detections of CO emission in $z$$>$1 galaxies. {\em
    Left}:\ Total number of detections (red) and detections per year
  (yellow) since the initial detection in 1991/1992. {\em
    Right}:\ Detections as of 2011 as a function of redshift, and color
  encoded by galaxy type (see Sect.~3; Riechers et al., in prep.).}
   \label{f1}
\vspace*{-3.5mm}

\end{figure}
%%%%%%%%%%%%%%%%%%%%%%%%%%%%%%%%%%%%%%%%%%%%%%%%%%%%%%

A number of different high-$z$ galaxy populations have been studied in
CO emission to date. These populations can be split into eight (at
least partially overlapping) categories according to their selection,
but fall into two main categories:\ (sub)millimeter-luminous and
(sub)millimeter-faint galaxies (see Fig.~1; Riechers et al., in
prep.). Historically, (sub)millimeter brightness was considered a
prerequisite for CO searches. However, with the improvement in
sensitivity and bandwidth of interferometers used for CO searches
since $\sim$3\,years ago, this restriction could be overcome.

The (sub)millimeter-luminous systems are typically either directly
selected in the observed-frame (sub)millimeter continuum
(submillimeter galaxies) or pre-selected through other diagnostics
(typically brightness of an active galactic nucleus (AGN) in the
optical or radio, i.e., quasars and radio galaxies) and then detected
in the (sub)millimeter continuum before they are followed up in CO
emission.  The (sub)millimeter-faint systems are pre-selected through
other star formation diagnostics in the optical/near-infrared,
mid-infrared and radio continuum (or a combination), and classified by
this selection (massive gas-rich star-forming galaxies, lensed
Lyman-break galaxies, extremely red objects, 24\,$\mu$m-selected
galaxies, radio-selected star-forming galaxies). Some of these samples
are specifically selected against being (sub)millimeter-luminous
(usually to probe galaxies with different dust and gas properties than
SMGs), while others may contain few systems that can be classified as
SMGs.  These samples are, however, not systematically selected against
each other, and thus contain at least partial overlap.

%%%%%%%%%%%%%%%%%%%%%%%%%%%%%%%%%%%%%%%%%%%%%%%%%%%%%%%%%
%%%% Fig.2: AzTEC-3 & J1148
%%%%%%%%%%%%%%%%%%%%%%%%%%%%%%%%%%%%%%%%%%%%%%%%%%%%%%%%%

\begin{figure}[h!]

\vspace*{-4mm}
      \begin{center}
%\hspace*{-60mm}
\includegraphics[width=12.5cm,angle=-0]{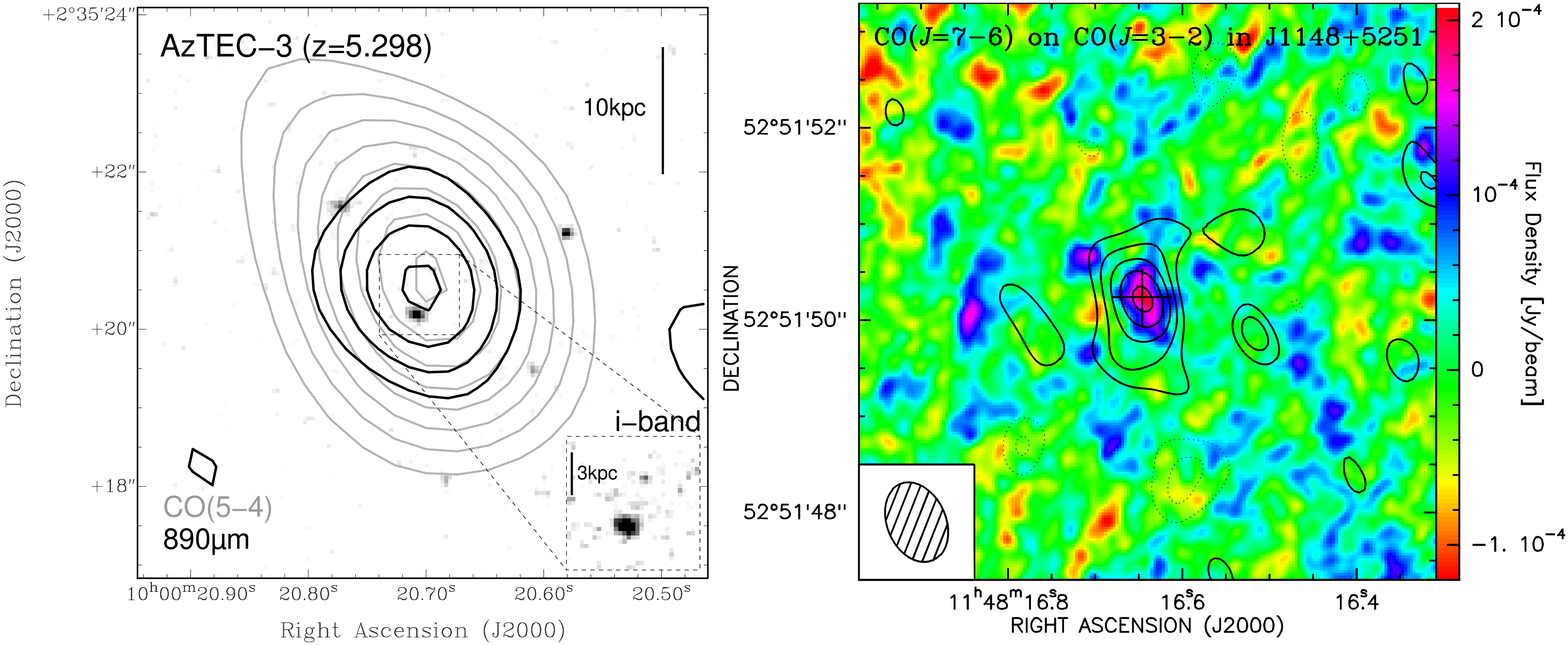}
      \end{center}
\vspace*{-8mm}

\caption{Highest redshift detections of CO emission in SMGs and
  quasars. {\em Left}:\ CO($J$=5$\to$4) emission in the $z$=5.298 SMG
  AzTEC-3 (PdBI; gray contours), overlaid on a HST $i$-band image. The
  black contours indicate the 890\,$\mu$m (rest-frame 140\,$\mu$m)
  continuum emission, as measured by the SMA. The inset shows a
  zoom-in of the region where the CO and 890\,$\mu$m emission peak
  (Riechers et al.\ 2010c). {\em Right}:\ CO($J$=7$\to$6) image of the
  $z$=6.42 quasar J1148+5251 (PdBI contours), overlaid on a 0.3$''$
  resolution CO($J$=3$\to$2) image obtained with the VLA. The
  molecular gas in the host galaxy of this quasar is distributed over
  5\,kpc scales (Walter et al.\ 2004; Riechers et al.\ 2009a). }
   \label{f2}
\vspace*{-3.6mm}
%\vspace*{-3.6mm}

\end{figure}
%%%%%%%%%%%%%%%%%%%%%%%%%%%%%%%%%%%%%%%%%%%%%%%%%%%%%%

%\subsection{
{\bf Quasar Host Galaxies (QSOs)}:\ Quasars were the first high-$z$
galaxies detected in CO emission (e.g., Brown \& Vanden Bout 1991;
Solomon et al.\ 1992; Barvainis et al.\ 1994; Ohta et al.\ 1996; Omont
et al.\ 1996). To date, the host galaxies of 34 $z$$>$1 quasars were
detected in CO emission, 14 of which are gravitationally
lensed. Initial detections are typically obtained in mid-$J$ CO lines
redshifted to 3\,mm.  All of these quasars are far-infrared-luminous,
i.e., detected at observed-frame (sub)millimeter wavelengths (see
Riechers 2011 for a recent summary).  Almost half of the detections
are at $z$$>$4, reaching out to $z$=6.42 (Fig.~2). This makes quasars
the best-studied population in CO emission at very high $z$. Also,
they were probed in a wide range of CO transitions, reaching from CO
$J$=1$\to$0 up to $J$=11$\to$10 (e.g., Weiss et al.\ 2007). They have
molecular gas masses in the range of
$\sim$0.1--20$\times$10$^{10}$\,$M_\odot$. The faint end
($\lesssim$10$^{10}$\,$M_\odot$) can currently only be probed with the
help of gravitational lensing (e.g., Riechers 2011).

%\subsection
{\bf Submillimeter Galaxies (SMGs)}:\
Since 1998, 34 (sub)millimeter galaxies were detected in CO emission,
16 of which are considered to be gravitationally lensed (e.g., Frayer
et al.\ 1998; Solomon \& Vanden Bout 2005; Engel et al.\ 2010; Lupu et
al.\ 2011). CO is detected out to the most distant SMG currently known
at $z$=5.3 (Fig.~2; Riechers et al.\ 2010c; Capak et al.\ 2011).
Initial detections are typically obtained in mid-$J$ CO lines
redshifted to 3\,mm.  Historically, most SMGs were selected at
850\,$\mu$m or 1.1/1.2\,mm, but with the advent of Herschel, the
selection has at least partially shifted to shorter wavelengths
(250--500\,$\mu$m, e.g., Negrello et al.\ 2010). SMGs have molecular
gas masses in the range of $\sim$0.3--15$\times$10$^{10}$\,$M_\odot$,
the least massive of which are typically strongly lensed (e.g., Sheth
et al.\ 2004; Kneib et al.\ 2005).

%\subsection{
{\bf Massive Gas-Rich Star-Forming Galaxies (SFGs)}:\
Daddi et al.\ (2008, 2010a) and Tacconi et al.\ (2010) have observed a
population of high-$z$ massive, gas-rich star-forming galaxies in CO
emission that are undetected at (sub)millimeter wavelengths. The six
galaxies by Daddi et al.\ are selected through the so-called BzK
technique (a two color selection around the redshifted
4000\,\AA\ break) to be massive, star-forming galaxies. These galaxies
are at $z$=1.4--1.6, and CO $J$=2$\to$1 was detected in all of
them. Tacconi et al.\ selected a sample of 19 massive star-forming
galaxies at $z$-1.1--2.4 in the ultraviolet/optical (overlapping with
massive Lyman-break galaxies), of which they detect 16 in CO
$J$=3$\to$2 emission. The parent populations of these galaxies are
more than an order of magnitude more common than SMGs, reaching
projected space densities of $>$1\,arcmin$^2$ (e.g., Daddi et
al.\ 2007).  For the combined sample of 22 detections, gas masses of
$\sim$2--50$\times$10$^{10}$\,$M_\odot$ are found.\footnote{In
  contrast to all other samples discussed here, a conversion factor of
  $M_{\rm gas}$/$L'_{\rm CO}$=$\alpha_{\rm
    CO}$=3.2--3.6\,$M_\odot$\,(K\,km\,s$^{-1}$\,pc$^2$)$^{-1}$ is used
  here, rather than 0.8\,$M_\odot$\,(K\,km\,s$^{-1}$\,pc$^2$)$^{-1}$.}

%\subsection
{\bf Star-Forming Radio-selected Galaxies (SFRGs)}:\
Chapman et al.\ (2008) and Casey et al.\ (2009) have observed a sample
of (sub)millimeter-undetected star-forming galaxies at $z$=1.3--2.2
selected through their 20\,cm radio continuum fluxes in CO emission.
They detect 8/14 galaxies in CO $J$=2$\to$1, $J$=3$\to$2, or
$J$=4$\to$3, emission, finding typical molecular gas masses of
8$\times$10$^{9}$\,$M_\odot$ (including two BzK-selected galaxies from
Daddi et al.\ 2008 into their sample). The selection of these galaxies
appears to overlap with the BzK selection.

%\subsection
{\bf 24\,$\mu$m-selected Galaxies (MIPS Galaxies)}:\
Yan et al.\ (2010) have observed CO $J$=2$\to$1 or $J$=3$\to$2
emission toward nine ultra-luminous infrared galaxies (ULIRGs) at
$z$=1.6--2.5 selected by their high 24\,$\mu$m luminosities
(24\,$\mu$m fluxes of 1.1--1.5\,mJy as measured with Spitzer/MIPS).
These galaxies are typically undetected at 1.2\,mm down to
$\sim$1\,mJy, suggesting that they trace a different high-$z$ ULIRG
population than (sub)millimeter-selected galaxies. These galaxies are
typically mergers, and their rest-frame mid-infrared luminosities
appear to be dominated by dust-obscured AGN. Yan et al.\ detect 8/9
galaxies, suggesting an average molecular gas mass of
1.7$\times$10$^{10}$\,$M_\odot$. Iono et al.\ (2006) report a
detection of CO $J$=2$\to$1 and $J$=3$\to$2 emission from another
24\,$\mu$m-selected galaxy, which however appears to be a strongly
lensed SMG (and thus, is counted as such here).

%\subsection
{\bf Lensed Lyman-break Galaxies (LBGs)}:\
With current instrumentation, it is not possible to detect `typical'
Lyman-break Galaxies (LBGs) in molecular line emission. However, aided
by strong gravitational lensing, Baker et al.\ (2004) and Coppin et
al.\ (2007) detected CO($J$=3$\to$2) emission toward two young
$z$$\sim$3 LBGs that are lensing-magnified by factors of $\mu_{\rm
  L}$=$\sim$30. One of these galaxies is weakly detected in continuum
emission at 1.2\,mm, but even with such strong lensing magnification
not at levels comparable to SMGs.  Through subsequent CO($J$=1$\to$0)
observations, Riechers et al.\ (2010a) find their (lensing-corrected)
total molecular gas masses to be 5--9$\times$10$^{8}$\,$M_\odot$,
which are the lowest currently observed at high $z$.

%\subsection
{\bf Radio Galaxies (RGs)}:\
Six galaxies at $z$$\sim$2.4--3.8 selected through powerful radio AGN
were detected in molecular line emission (e.g., Scoville et al.\ 1997;
Papadopoulos et al.\ 2000; De Breuck et al.\ 2003a, 2003b, 2005; Greve
et al.\ 2004).  These six galaxies also show strong (sub)millimeter
continuum emission. They have typical molecular gas masses of
5--10$\times$10$^{10}$\,$M_\odot$.

%\subsection
{\bf Extremely Red Objects (EROs)}:\ Andreani et al.\ (2000) and Greve
et al.\ (2003) have detected CO emission from an Extremely Red Object
(i.e., with very red optical/infrared colors of $R$--$K$$>$6) at
$z$=1.4, finding a total molecular gas mass of
6$\times$10$^{10}$\,$M_\odot$. Despite its selection in the
optical/infrared, this galaxy appears to show properties that are
similar to SMGs (and is, indeed, detected at submillimeter wavelengths
in the continuum), but possibly with a lower-than-average CO
excitation within this population.

\section{CO Excitation}

%%%%%%%%%%%%%%%%%%%%%%%%%%%%%%%%%%%%%%%%%%%%%%%%%%%%%%%%%
%%%% Fig.3: CO excitation plot
%%%%%%%%%%%%%%%%%%%%%%%%%%%%%%%%%%%%%%%%%%%%%%%%%%%%%%%%%

\begin{figure}[h!]

\vspace*{-5mm}
      \begin{center}
%\hspace*{-60mm}
\includegraphics[width=12.5cm,angle=-0]{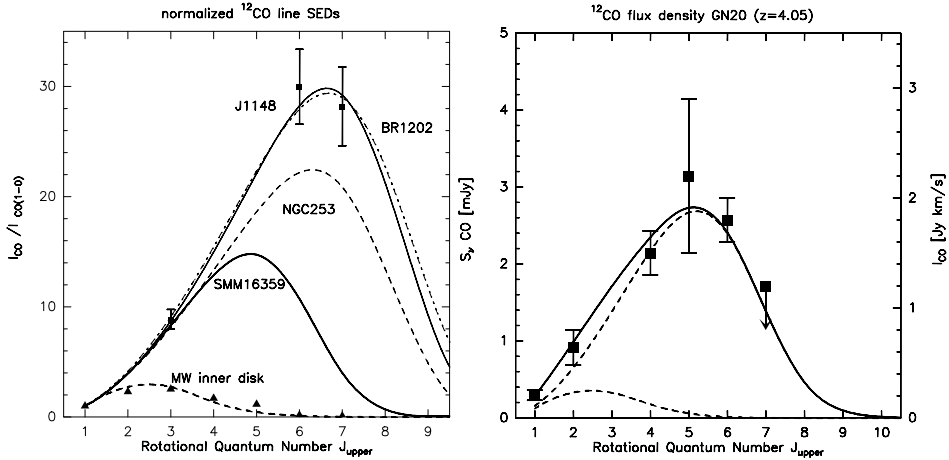}
      \end{center}
\vspace*{-8mm}

\caption{CO excitation diagrams and LVG models (lines) for high-$z$
  galaxies. {\em Left}:\ Comparison of CO line ladders of the $z$=6.42
  quasar J1148+5251 (squares and solid line) and the $z$=4.69 quasar
  BR\,1202-0725 (dash-dotted line) to the nucleus of the nearby
  starburst galaxy NGC\,253 (dashed line), the $z$=2.52 SMG
  J16359+6612 (thick solid line), and the inner disk of the Milky Way
  (triangles and thick dashed line; Riechers et al.\ 2009a). {\em
    Right}:\ CO excitation of the $z$=4.05 SMG GN20. The dashed lines
  indicate the two gas components required to fit the observed CO
  excitation, and the solid line shows the sum of both components
  (Carilli et al.\ 2010).}
   \label{f3}
%\vspace*{-5.5mm}
\vspace*{-3.6mm}

\end{figure}
%%%%%%%%%%%%%%%%%%%%%%%%%%%%%%%%%%%%%%%%%%%%%%%%%%%%%%

In lack of spatial resolution in most studies of CO emission out to
the largest cosmic distances, one of the most insightful diagnostics
is to study the CO excitation of high-$z$ galaxies. In an idealized
case of thermal equilibrium, all rotational transitions of CO in a
galaxy would exhibit the same brightness temperature and line
luminosity $L'_{\rm CO}$, which implies that their peak flux densities
increase toward higher-$J$ lines, scaling with $\nu^2_{\rm
  line}$. However, the energy requirements to keep higher-$J$ CO
lines in thermal equilibrium get increasingly higher (which can be
expressed as higher critical densities and excitation temperatures),
making it increasingly more difficult to maintain equilibrium. Thus,
by modeling the intensity of different rotational levels of CO
relative to equilibrium, it is possible to constrain the physical
properties of the gas (in particular its density and kinetic
temperature). In lack of detailed constraints on the gas distribution
and local radiation field, the most common approach is to model the
collisional excitation of CO using the large velocity gradient (LVG)
approximation (e.g., Scoville \& Solomon 1974; Goldreich \& Kwan
1974). This method utilizes an escape probability formalism (i.e.,
photons produced locally can only be absorbed locally) resulting from
a strong velocity gradient, which helps to minimize the number of free
parameters in models of the collisional line excitation. 

By measuring the fluxes of multiple CO transitions in high-$z$
galaxies, LVG models allow for a comparison of the gas properties of
different high-$z$ galaxy populations (see Fig.~3 for examples). High
redshift quasars appear to be dominated by high-excitation gas
components, where the line peak fluxes typically start to decrease in
$J$$>$6 transitions. This yields characteristic gas kinetic
temperatures of $T_{\rm kin}$=50\,K and characteristic gas densities
$n_{\rm gas}$ of few times 10$^4$\,cm$^{-3}$ (e.g., Riechers et
al.\ 2006a, 2009a), comparable to what is found in the nuclei of
nearby ULIRGs. More extreme examples such as the $z$=3.911 quasar
APM\,08279+5255 are observed, but for such systems, significant
contributions from radiative excitation from the AGN and/or
star-forming regions remain a possibility (e.g., Wei\ss\ et
al.\ 2007). In contrast, SFGs appear to be dominated by colder, less
dense gas components in which the line peak fluxes typically start to
decrease in $J$$>$3 transitions. This yields characteristic gas
kinetic temperatures of $T_{\rm kin}$=25\,K and characteristic gas
densities $n_{\rm gas}$ of few times 10$^3$\,cm$^{-3}$ (Dannerbauer et
al.\ 2009; Aravena et al.\ 2010), comparable to what is found in
nearby spiral/disk galaxies and the Milky Way. Given the integrated
star formation rates of SFGs, the presence of some denser, higher
excitation gas is expected, but the relative strength of such
components is not constrained well at present. SMGs appear to contain
a mix of dense, high excitation gas (even though somewhat lower than
in quasars on average, with line peak fluxes starting to decrease in
$J$$>$5 transitions) and low excitation gas as found in SFGs (Fig.~3,
right; e.g., Carilli et al.\ 2010; Harris et al.\ 2010; Riechers et
al.\ 2010c; Ivison et al.\ 2011). The high-exitation gas component in
SMGs typically dominates the line fluxes in $J$$\geq$2 lines. However,
due to a potential difference in $\alpha_{\rm CO}$=$M_{\rm
  gas}$/$L'_{\rm CO}$, this does not imply that this component
dominates the gas mass (e.g., Carilli et al.\ 2010; Riechers et
al.\ 2010c). Only limited information is available on the CO
excitation of young, lensed LBGs to date, but their excitation appears
consistent with the somewhat intermediate levels found in nearby
luminous infrared galaxies (Riechers et al.\ 2010a).

\section{Luminosity Relations and the Star Formation Law}

The CO luminosity $L'_{\rm CO}$ is commonly considered to be a measure
for the total molecular gas mass $M_{\rm gas}$ in a galaxy, while the
(far-) infrared luminosity $L_{\rm FIR}$ is considered to be a measure
for the star formation rate (SFR; e.g., Solomon \& Vanden Bout 2005).
Thus, the relation between $L'_{\rm CO}$ or $M_{\rm gas}$ and $L_{\rm
  FIR}$ or SFR may be considered a spatially integrated version of the
Schmidt--Kennicutt `star formation law' between gas surface density
and star formation rate surface density (e.g., Schmidt 1959; Kennicutt
1998).

Even in galaxies with luminous AGN, $L_{\rm FIR}$ is commonly used as
a proxy for the star formation rate in the host galaxy. In principle,
both the AGN and star formation can heat the dust that gives rise to
the continuum flux observed in the FIR, but the characteristic dust
temperatures of AGN heating are typically a factor of a few higher
than those of dust heating by young stars. Thus, the warm dust in
AGN-starburst systems observed in the rest-frame FIR is commonly
thought to be dominated by heating within the host galaxy, in
particular in the most intense, dust-enshrouded starbursts. If,
however, $L_{\rm FIR}$ were to be dominated by the AGN, one would
expect an elevated $L_{\rm FIR}$ for such galaxies in the $L'_{\rm
  CO}$--$L_{\rm FIR}$ relation. In Figure 4 (right panel), a
comparison of the $L'_{\rm CO}$--$L_{\rm FIR}$ relation for nearby and
high-$z$ quasars to nearby galaxies, ULIRGs and SMGs without dominant
AGN is shown (Riechers 2011). For galaxies with low $L'_{\rm CO}$,
there is an indication for an excess in $L_{\rm FIR}$ for quasars
relative to other systems; however, there is no evidence for such a
trend at high $L'_{\rm CO}$. This may suggest that, in systems with
relatively low gas and dust content, the AGN contributes significantly
to $L_{\rm FIR}$, but not in systems with high gas and dust
content. Thus, for the massive high-$z$ systems, $L_{\rm FIR}$ appears
to be a good proxy for the SFR even in quasars (Riechers 2011).

%%%%%%%%%%%%%%%%%%%%%%%%%%%%%%%%%%%%%%%%%%%%%%%%%%%%%%%%%
%%%% Fig.4: SF laws
%%%%%%%%%%%%%%%%%%%%%%%%%%%%%%%%%%%%%%%%%%%%%%%%%%%%%%%%%

\begin{figure}[h!]

\vspace*{-4mm}
      \begin{center}
%\hspace*{-60mm}
\includegraphics[width=12.5cm,angle=-0]{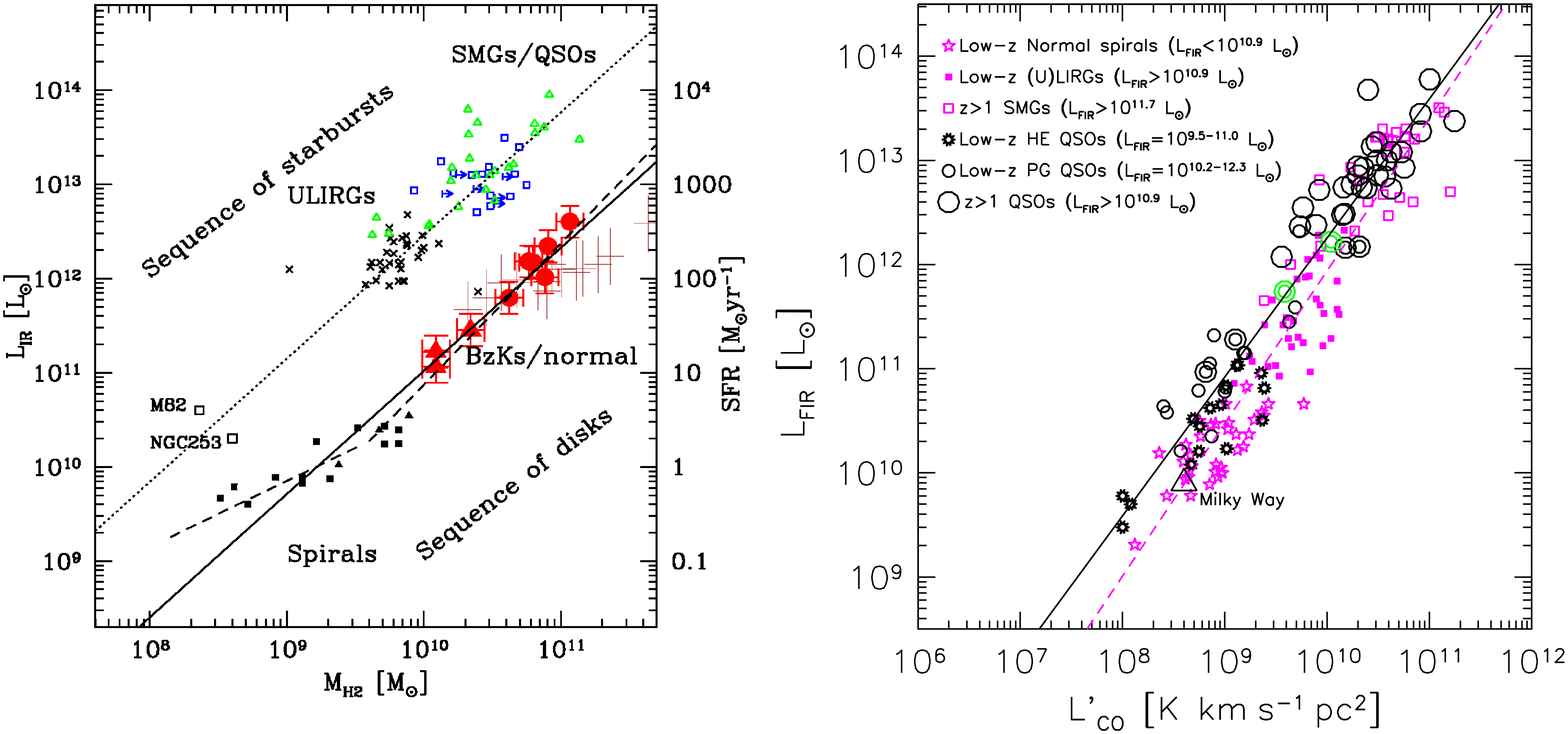}
      \end{center}
\vspace*{-8mm}

\caption{Comparison of CO luminosity/gas mass with (far-) infrared
  luminosity as a tracer of the star formation rate in low- and
  high-$z$ galaxies. {\em Left}:\ $M_{\rm gas}$--$L_{\rm IR}$ relation
  for low-$z$ spirals, starbursts and ULIRGs, and high-$z$ SFGs
  (``BzKs/normal''), SMGs, and quasars. The solid line represents a
  fit to spiral galaxies and SFGs, and the dotted line shows the same
  trend offset by 1.1dex in $L_{\rm IR}$ (which is consistent with the
  starburst galaxies, ULIRGs, SMGs and quasars). These two lines may
  represent two sequences for disk and starburst galaxies, with the
  offset being due to different dynamical timescales for star
  formation (e.g., Daddi et al.\ 2010b). {\em Right}:\ $L'_{\rm
    CO}$--$L_{\rm FIR}$ relation for quasars at low and high $z$
  (fitted by the solid line), and galaxies without dominant AGN at low
  $z$ (spirals and (U)LIRGs) and at high $z$ (SMGs; fitted by the
  dashed line). At high $L'_{\rm CO}$, galaxies with and without
  dominant AGN statistically occupy the same region, while at low
  $L'_{\rm CO}$, there is tentative evidence for an offset of quasars
  toward higher $L_{\rm FIR}$. This may indicate that AGN heating
  contributes significantly to $L_{\rm FIR}$ at low $L'_{\rm CO}$, but
  not at high $L'_{\rm CO}$ (Riechers 2011).}
   \label{f4}
%\vspace*{-6.6mm}
\vspace*{-3.6mm}

\end{figure}
%%%%%%%%%%%%%%%%%%%%%%%%%%%%%%%%%%%%%%%%%%%%%%%%%%%%%%

More fundamentally, the question occurs if star formation progresses
the same way in all types of galaxies. In disk-like spiral galaxies
like the Milky Way, star formation occurs in molecular clouds with
compact, dense cores, confined by self gravity (e.g., Solomon et
al.\ 1987). In mergers, such as the Antennae (NGC\,4038/39), star
formation appears to peak on relatively large scales in the dense
overlap region between the merging galaxies, leading to the formation
of so-called super star clusters (e.g., Wilson et al.\ 2003). In the
nuclei of ULIRGs, the most extreme nearby starbursts, star formation
appears to occur in a dense, intercloud medium, bound by the potential
of the galaxy, rather than virialized clouds (e.g., Downes \& Solomon
1998). As a first consequence, the conversion factor from observed CO
luminosity to gas mass $M_{\rm gas}$/$L'_{\rm CO}$=$\alpha_{\rm CO}$
is different for these different kind of galaxies (e.g., $\alpha_{\rm
  CO}$=0.8\,$M_\odot$\,(K\,km\,s$^{-1}$\,pc$^2$)$^{-1}$ is typically
used for ULIRGs, while $\alpha_{\rm
  CO}$=4.6\,$M_\odot$\,(K\,km\,s$^{-1}$\,pc$^2$)$^{-1}$ is used for
the Milky Way). For ``ULIRG-like'' starbursts at high $z$ such as
quasars, SMGs, and 24\,$\mu$m-selected galaxies (many of which show
merger-like characteristics, e.g., Fig.~5), an ULIRG conversion factor
is commonly adopted (e.g., Solomon \& Vanden Bout 2005). For SFGs,
$\alpha_{\rm
  CO}$=3.2--3.6\,$M_\odot$\,(K\,km\,s$^{-1}$\,pc$^2$)$^{-1}$ is
typically used, comparable to those in nearby disk galaxies (e.g.,
Daddi et al.\ 2010a; Tacconi et al.\ 2010).

%%%%%%%%%%%%%%%%%%%%%%%%%%%%%%%%%%%%%%%%%%%%%%%%%%%%%%%%%
%%%% Fig.5: Velocity maps
%%%%%%%%%%%%%%%%%%%%%%%%%%%%%%%%%%%%%%%%%%%%%%%%%%%%%%%%%

\begin{figure}[h!]

\vspace*{-4mm}
      \begin{center}
%\hspace*{-60mm}
\includegraphics[width=8.5cm,angle=-0]{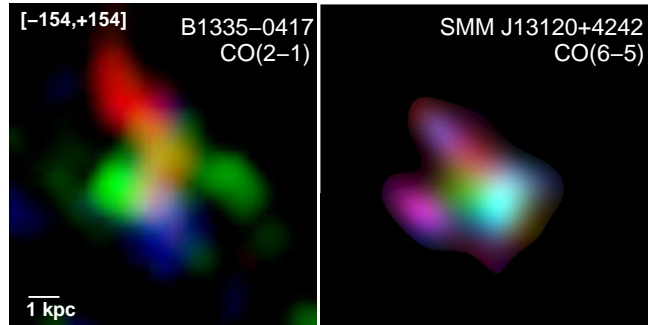}
      \end{center}
\vspace*{-8mm}
 
\caption{Velocity maps of the $z$=4.4 quasar host galaxy
  BRI\,1335-0417 (Riechers et al.\ 2008a), and the $z$=3.4 SMG
  J13120+4242 (Engel et al.\ 2010). BRI\,1335-0417 was observed at a
  resolution of $\sim$0.15$''$ ($\sim$1.0\,kpc), tapered to
  0.32$''$$\times$0.30$''$ here. J13120+4242 was observed at a
  resolution of 0.59$''$$\times$0.47$''$ ($\sim$4\,kpc). Both systems
  show a complex morphology and velocity structure inconsistent with a
  rotating disk, suggesting that they are major, gas-rich mergers with
  $>$5\,kpc scale gas reservoirs.}
   \label{f5}
%\vspace*{-6.6mm}
\vspace*{-3.6mm}

\end{figure}
%%%%%%%%%%%%%%%%%%%%%%%%%%%%%%%%%%%%%%%%%%%%%%%%%%%%%%

The differences between starburst and disk galaxies are also reflected
in the star formation law.  When comparing the $M_{\rm gas}$--$L_{\rm
  IR}$ relation for three largest CO-detected samples at high $z$,
i.e., quasars, SMGs, and SFGs to low redshift galaxies, two
interesting trends occur (using $\alpha_{\rm CO}$ as outlined above,
but note that the trends are visible independent of this
choice). First, SFGs extend the relation found for nearby spiral
galaxies to higher $M_{\rm gas}$. Second, quasars and SMGs extend the
trend found for the most intense nearby starbursts and ULIRGs to
higher $M_{\rm gas}$. Both trends agree with the same slope, but are
offset by 1.1dex in $L_{\rm IR}$. Daddi et al.\ (2010b) interpret this
as evidence for two sequences in this relation for disk galaxies and
starbursts, which occur due to the different dynamical timescales of
star formation in these systems (Fig.~4, left; see also Genzel et
al.\ 2010).

\section{Detectability of CO Emission at High Redshift:\ Recent Developments}

Until recently, all CO detections at $z$$>$1 were obtained based on
precise, spectroscopic redshifts from optical/infrared diagnostics,
biasing studies toward those systems that are bright at such
wavelengths. This is due to the fact that, until recently,
radio/millimeter-wave telescopes had very narrow fractional bandwidths
that covered less or little more than the (typically few
100\,km\,s$^{-1}$ wide) CO emission lines. However, with the recent
advent of so-called ``Z-machines'' like EMIR on the 30\,m telescope
(8\% fractional bandwidth), the Zpectrometer on the Green Bank
Telescope (GBT; 34\% fractional bandwidth at 1\,cm) and Z-spec on the
Caltech Submillimeter Observatory (CSO; 46\% fractional bandwidth at
1\,mm -- instrument is currently on the Atacama Pathfinder Experiment
(APEX)) and wide-band correlators on millimeter interferometers, in
particular the Plateau de Bure Interferometer (PdBI) and the Combined
Array for Research in Millimeter-wave Astronomy (CARMA; 4\%--8\%
fractional bandwidth at 3\,mm), this situation has drastically
changed. The first optically faint high-$z$ galaxy redshift to be
identified directly through CO emission was the $z$=4.05 SMG GN20
(Daddi et al.\ 2009), but this discovery was serendipitous. Wei\ss\ et
al.\ (2009) have identified a lensed $z$=2.93 SMG through a CO search
with EMIR on the 30\,m. Lupu et al.\ (2011) and Frayer et al.\ (2011)
have identified five 1.5$<$$z$$<$3.0 lensed SMGs with Z-spec and the
Zpectrometer on the CSO and GBT. Scanning the 3\,mm band with CARMA,
Riechers (2011) has identified a lensed type-2 quasar at $z$=2.06 in
CO. The 3\,mm band covers a least one CO transition at virtually any
redshift $<$0.44 and $>$1.0, making it a particularly suitable band
for CO redshift searches.  These studies show that, with the recent
upgrade in instrumentation (and even more so with ALMA and the fully
upgraded EVLA in the future), it has now become possible to directly
identify galaxy redshifts through CO emission, liberating studies of
molecular gas at high $z$ from one of their main biases. By targeting
strongly lensed sources, these studies greatly increase the number of
galaxies with detected CO emission. Also, they offer a great
demonstration of what studies ALMA will enable in 1--2 orders of
magnitude fainter systems.

%%%%%%%%%%%%%%%%%%%%%%%%%%%%%%%%%%%%%%%%%%%%%%%%%%%%%%%%%
%%%% Fig.6: CO searches
%%%%%%%%%%%%%%%%%%%%%%%%%%%%%%%%%%%%%%%%%%%%%%%%%%%%%%%%%

\begin{figure}[h!]

\vspace*{-5mm}
      \begin{center}
%\hspace*{-60mm}
\includegraphics[width=12.5cm,angle=-0]{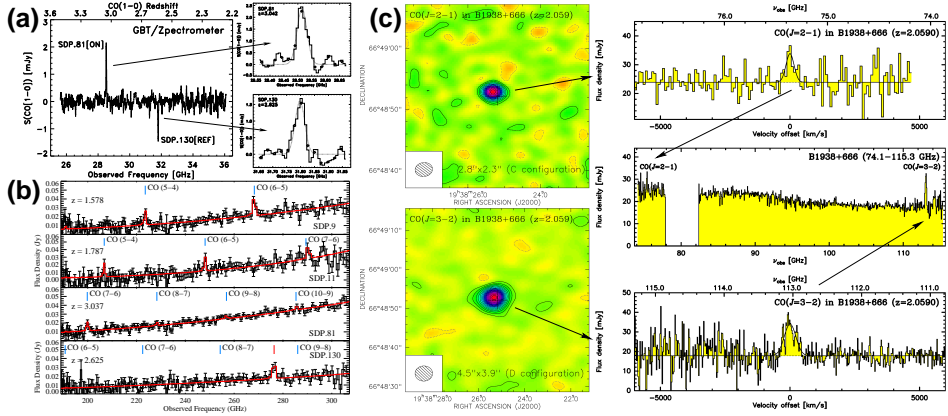}
      \end{center}
\vspace*{-8mm}

\caption{``Blind'' CO searches toward high-$z$ galaxies. {\em
    a}:\ Detection of CO($J$=1$\to$0) emission toward two lensed
  $z$$>$2 SMGs with the GBT/Zpectrometer, redshifted to $\sim$1\,cm
  wavelength (Frayer et al.\ 2011). The instrument position-switched
  between both sources for calibration purposes, making one of the
  lines artificially appear as negative in the difference
  spectrum. {\em b}:\ Detection of high-$J$ CO emission toward four
  lensed $z$$>$1.5 SMGs with CSO/Z-spec, redshifted to $\sim$1\,mm
  wavelength (Lupu et al.\ 2011), two of which are the same sources as
  in panel {\em a}. {\em c}:\ Detection of CO $J$=2$\to$1 and 3$\to$2
  emission toward a $z$=2.06 type-2 quasar, scanning the 3\,mm band
  with CARMA (Riechers 2011). These are some of the first examples of
  galaxies that were identified through CO redshifts with wideband
  spectrometers on single-dish telescopes and wide-band correlators on
  millimeter interferometers.}
   \label{f5}
%\vspace*{-6.6mm}
\vspace*{-3.6mm}

\end{figure}
%%%%%%%%%%%%%%%%%%%%%%%%%%%%%%%%%%%%%%%%%%%%%%%%%%%%%%

\section{Implications for Future Systematic Studies of Molecular Gas in the Early Universe}

Based on studies of molecular gas in selected high-$z$ galaxies, we
have learned that we can characterize different galaxy populations in
the early universe based on the properties of their molecular
interstellar medium content. Also, recent technological improvements
have yielded direct ``blind'' redshift identifications of high-$z$
galaxies through CO line emission. Moreover, studies of CO emission in
FIR-faint galaxies have shown that the projected density of
CO-luminous high-$z$ galaxies likely exceeds 1\,arcmin$^{-2}$, which
is comparable to the field-of-view of ALMA at 3\,mm, and a fraction of
the field-of-view of the EVLA at 1\,cm. In other words, sufficiently
deep studies that probe sufficient cosmic volume (by covering a large
range in redshift for CO emission lines) with these instrument may
yield detections of CO emission in high-$z$ galaxies pointing at any
position in the sky, despite their limited field-of-view. 

This suggests that studies of redshifted CO emission comparable to
optical ``Deep Fields'' (i.e., blindly targeted
$\gtrsim$10\,arcmin$^2$ size fields studied to faint levels) will
become feasible with the next generation of radio/(sub)millimeter
telescopes, and are a critical complement to CO intensity mapping
studies at $>$5$'$ resolution that are aimed at studying large scale
structure and that will place valuable constraints on cosmic
reionization (e.g., Gong et al.\ 2011; Carilli 2011; Bowman et al., in
prep.). At optical/infrared wavelengths, Deep Field studies have
provided critical constraints on the SFHU and the stellar mass density
of the universe. Comparable studies in molecular line emission have
the potential to provide comparable constraints on the prerequisite
material for star formation, i.e., the molecular gas mass density of
the universe (which is linked to the SFHU through the star formation
law, and is the material available for stellar mass assembly in
galaxies). Such a Deep Field study of the molecular gas mass density
of the universe thus has the potential to substantially further our
understanding of galaxy formation throughout cosmic times.

\acknowledgements I would like to thank my collaborators on a number
of studies related to this subject, in particular Manuel Aravena,
Frank Bertoldi, Peter Capak, Chris Carilli, Ashanta Cooray, Pierre
Cox, Emanuele Daddi, Helmut Dannerbauer, Roberto Neri, Nick Scoville,
Jeff Wagg, Fabian Walter, Ran Wang, and Axel Wei\ss. I acknowledge
support from NASA through Hubble Fellowship grant HST-HF-51235.01
awarded by STScI, operated by AURA for NASA, under contract NAS
5-26555.\\[-6mm]

%\bibliography{aspauthor}

\end{document}